\documentclass[a4paper,conference,10pt]{IEEEtran}
\IEEEoverridecommandlockouts
\usepackage{cite}
\usepackage{amsmath,amssymb,amsfonts}
\usepackage{algorithmic}
\usepackage{graphicx}
\usepackage{textcomp}
\usepackage{xcolor}
\def\BibTeX{{\rm B\kern-.05em{\sc i\kern-.025em b}\kern-.08em
    T\kern-.1667em\lower.7ex\hbox{E}\kern-.125emX}}

\usepackage[nowarn,acronyms,nonumberlist,nopostdot,nomain,nogroupskip]{glossaries}

\usepackage{caption}
\usepackage{subcaption}
\usepackage[inline]{enumitem}

\newacronym{sinr}{SINR}{Signal to Interference plus Noise Ratio}
\newacronym{ar}{AR}{Augmented Reality}
\newacronym{vr}{VR}{Virtual Reality}
\newacronym{dpc}{DPC}{Dirty Paper Coding}
\newacronym{sc}{SC}{Superposition Coding}
\newacronym{noma}{NOMA}{Non-Orthogonal Multiple Access}
\newacronym{sic}{SIC}{Successive Interference Cancellation}
\newacronym{oma}{OMA}{Orthogonal Multiple Access Schemes}
\newacronym{rsma}{RSMA}{Rate Splitting Multiple Access}
\newacronym{rs}{RS}{Rate Splitting}
\newacronym{hrs}{HRS}{Hierarchical Rate Splitting}
\newacronym{csi}{CSI}{Channel State Information}
\newacronym{csit}{CSIT}{Channel State Information at the Transmitter}
\newacronym{uca}{UCA}{Uniform Circular Array}
\newacronym{ula}{ULA}{Uniform Linear Array}
\newacronym{nn}{NN}{Neural Network}
\newacronym{snr}{SNR}{Signal-to-Noise ratio}

\newacronym{rzf}{RZF}{Regularized Zero Forcing}

\newacronym{mbf}{MBF}{Matched Beamforming}

\newacronym{doa}{DoA}{Direction-of-Arrival}

\usepackage{pgfplots}
\usepackage{svg}  
\usetikzlibrary{plotmarks}
\usetikzlibrary{arrows.meta}
\usepgfplotslibrary{patchplots}

\usepackage{svg}

\pgfplotsset{compat=1.17}

\begin{document}


\title{
User Clustering  for Rate Splitting using\\ Machine Learning
\thanks{
The authors' affiliations and emails are as follows:
\\$^1$ISPIC, Centre Tecnològic Telecomunicacions Catalunya, Barcelona, Spain. Emails: \{rpereira, xmestre\}@cttc.es\\
$^2$Department of Information Engineering, University of Padova, Padova, Italy. Emails: \{deshpande, zanella\}@dei.unipd.it \\
$^3$Department of Electronic Systems, Aalborg University, Aalborg, Denmark. Emails: \{cjvr, edc, petarp\}@es.aau.dk \\
$^4$Software Radio Systems, Barcelona, Spain, Email:
 david.gregoratti@ieee.org
 \\
 This work has been partially funded by the European Commission under the Windmill project (contract 813999) and the Spanish government under the Aristides project (RTI2018-099722-B-I00).
}
}

\author{Roberto Pereira$^{1}$, Anay Ajit Deshpande$^{2}$, Cristian J. Vaca-Rubio$^3$, \\Xavier Mestre$^1$, Andrea Zanella$^2$, David Gregoratti{$^4$}, Elisabeth de Carvalho$^3$, Petar Popovski$^3$  \\
}

\maketitle

\begin{abstract}
Hierarchical Rate Splitting (HRS) schemes  proposed in recent years have shown to provide significant improvements in exploiting spatial diversity in wireless networks and provide high throughput for all users while minimising interference among them. Hence, one of the major challenges for such HRS schemes is the necessity to know the optimal clustering of these users based only on their Channel State Information (CSI). This clustering problem is known to be NP hard and, to deal with the unmanageable complexity of finding an optimal solution, in this work a scalable and much lighter clustering mechanism based on Neural Network (NN) is proposed. The accuracy and performance metrics show that the NN is able to learn and cluster the users based on the noisy channel response and is able to achieve a rate comparable to other more complex clustering schemes from the literature.

\end{abstract}


\section{Introduction}

Multi-antenna radio technologies have shown to enhance  spectral efficiency while ensuring connectivity to a large number of devices. Different encoding schemes such as \gls{dpc} have been designed to achieve the multi-antenna channel capacity \cite{yu2004sum}. However, due to the high computational complexity as well as the need for precise \gls{csi}, there has been much focus of research on sub-optimal solutions which combine \gls{sc} and spatial processing such as \gls{noma}\cite{goldsmith2005wireless}. Additionally, as these mechanisms tend to fully decode interference, the uncertainty over \gls{csi} directly affects interference cancellation among different users. Hence, the authors of \cite{mao2018rate} have recently proposed \gls{rsma} as a non-orthogonal transmission scheme that partially decodes interference and partially treats it as noise thus further improving multiplexing gains. For 1-layer \gls{rs}, the message intended to each user is divided  into a common ($s_c$) and private ($s_p$) parts encoded separately. In order for this transmission scheme to work, it is necessary to ensure that every user perfectly decodes the common message. This is often tackled by allocating a larger fraction of the total power to the common message. In the presence of a large number of receivers, this condition limits the total rate by the minimal common rate achieved in the whole system\footnote{This  happens regardless of the number of antennas at the transmitter. Instead, this is a consequence of power allocation to reduce interference among different users.}. Hence, in the presence of several users, the power assigned to each $s_p$ is reduced, leading to a degradation in communication rate. 

In these conditions, relying on multiple common streams (generalised rate splitting) leads to higher multiplexing gains, {but at the cost of high complexity at the decoder caused by the several layers of \gls{sic} \cite{mao2018rate}. To tackle the increasing complexity of generalised \gls{rs} while having small loss in multiplexing, the authors in \cite{rs_2rhs} consider a 2-layer \gls{hrs} transmission mechanism. In this scenario, users are considered to be divided into $G$ groups and  required to decode three messages: two common messages and a private message. One of the common messages (outer common - $s_{oc}$) is encoded using a codebook shared among all the user while the other one (inner common - $s_{ic,g}$) is encoded by a codebook share only among users in a specific group.} But when the groups are orthogonal, i.e. the users are sufficiently separated spatially, optimal communication happens when inter-group  and intra-group interference are reduced to a level that it can be completely distinguished from the intended signals.  








But, to minimise the interference and maximise the rate using \gls{hrs}, the Base Station (BS) is required to know what can be referred to as the optimal clustering scheme, i.e., the one that maximises the total communication rate. Unfortunately, finding this optimal clustering scheme is an NP hard problem which often requires an exhaustive search. Thus, it becomes extremely hard to come up with an optimisation mechanism that maximises the communication rate using HRS while also considering the clustering options as an optimisation variable. Hence, in this work, we propose a learning mechanism capable of directly learning  (or approximating) the optimal clustering option from the imperfect \Gls{csi}.

\section{System Model}
\label{sec:system_model}
Consider a downlink transmission scenario where $N$  single-antenna user equipment (UEs) receive messages from a base station (BS) over a spatially correlated Rayleigh-fading channel. We further assume this BS to be equipped with an antenna with $M$ isotropic antenna elements. 
Moreover, let these UEs be partitioned into $G \geq 1$ disjoint clusters. 
{So, the signal $\mathbf{y} \in \mathbb{C}^{N}$ received by all the users is given by}
\begin{equation}
    \mathbf{y} = \mathbf{H}^\mathrm{H} \mathbf{x} + \mathbf{n}
\end{equation}
where, $\mathbf{H} = [\mathbf{h}_{1}, \dots, \mathbf{h}_{N}]^\mathrm{T} \in \mathbb{C}^{M\times N}$ contains the stacked channels of all the $k~=\{1,\dots,N\}$ UEs, $\mathbf{n} \sim\mathcal{N}_\mathbb{C}(\mathbf{0},\mathbf{I}_{N})$ is an additive white Gaussian noise vector and $\mathbf{x} \in \mathbb{C}^M$ is the combined signal 
\begin{equation}\label{eq:hrs}
    \mathbf{x} = \sqrt{p_{oc}}\mathbf{w}_{oc}s_{oc} + 
    \sum_{g=1}^G \mathbf{B}_g \left( \sqrt{p_{ic,g}} \mathbf{w}_{ic,g} s_{ic,g} + \sqrt{p_{gk}}\mathbf{W}_{g}\mathbf{s}_{g}  \right)
\end{equation}
{where $p_{oc}$, $p_{ic,g}$ and $p_{gk}$ are the power allocated to the outer common message $s_{oc} \in \mathbb{C}$, inner common messages  $\mathbf{s}_{ic} \in \mathbb{C}^{G}$ and the private messages $\mathbf{s}_g \in \mathbb{C}^{N_g}$, respectively. $\mathbf{B}_g \in \mathbb{C}^{M \times b_g}$ is the group outer precoder designed from the $g$th group channel's second order statistics and dependent on the integer design parameters $b_g$ rank of the channel covariance matrix.}
By knowing the UEs that belong to the $g$th cluster, the matrix $\mathbf{H}_g = [\mathbf{h}_{g,1}, \dots, \mathbf{h}_{g,N_g}]^\mathrm{T} \in \mathbb{C}^{M\times N_g}$ contains the stacked channels of all the $N_g$ UEs that belong to the $g$th cluster. The downlink fading channel $\mathbf{h}_{g,k} \in  \mathbb{C}^{M}$ associated to the $k$th user of the $g$th class can be factored out  as
\begin{equation}
    \label{eq:channel}
    \mathbf{h}_{g,k} = \mathbf{R}_{g}^{\frac{1}{2}}\mathbf{g} = \mathbf{U}_{g}\mathbf{\Lambda}_{g}^{\frac{1}{2}}\mathbf{g}_k
\end{equation}
where $\mathbf{R}_g \in \mathbb{C}^{M\times M}$ is the channel correlation matrix, $\mathbf{U}_g \in \mathbb{C}^{M\times M}$ a  unitary matrix containing its eigenvectors,  $\mathbf{\Lambda}_g \in \mathbb{C}^{M\times M}$ a diagonal matrix with its associated eigenvalues and $\mathbf{g}_k \in \mathbb{C}^{M}$ has Gaussian independent and identically distributed (i.i.d.) entries with zero mean and unit variance which describe the complex path gains.



In principle, the covariance matrices are directly dependent on the angular response of the channels \cite{goldsmith2005wireless}. 
Unfortunately, in a more realistic environment, due to limited feedback, the BS only observes an imperfect estimation of the channel~(\ref{eq:channel}). Following~\cite{adhikary2013_jsdm_large}, we model this imperfection as the sum of a channel and a noise generated from the same subspace 
\begin{equation}
    \label{eq:channel:draw}
    \mathbf{\hat{h}}_{g,k} = \mathbf{U}_g \mathbf{\Lambda}_g^{\frac{1}{2}}\mathbf{\hat{g}}_k = \mathbf{U}_g \mathbf{\Lambda}_g^{\frac{1}{2}} \left(  \sqrt{1-\tau^2}\mathbf{g}_k + \tau\mathbf{z}_k  \right)
\end{equation}
where  $\mathbf{z}_k$ has i.i.d entries and $\tau \in [0,1]$ indicates the quality of the instantaneous channel. For instance,  $\tau=0$ leads to a perfect channel estimation, i.e., $\mathbf{\hat{h}}_{g,k} = \mathbf{R}_g^{\frac{1}{2}}\mathbf{g}_k$ while $\tau=1$ leads to an uncorrelated channel in the subspace spanned by $\mathbf{U}_g$, i.e.,  $\mathbf{\hat{h}}_{g,k} = \mathbf{R}_g^{\frac{1}{2}}\mathbf{z}_k$ for uncorrelated $\mathbf{g}_k$ and $\mathbf{z}_k$.

\subsection{HRS Transmission Mechanism }
{\gls{hrs} transmission design is defined based on the combined transmission signal $\mathbf{x}$ from \eqref{eq:hrs}. To determine the transmission signal $\mathbf{x}$, we obtain the precoder $\mathbf{B}_g$ following~\cite{adhikary2013_jsdm_large, rs_2rhs}, so that the group effective channel $\mathbf{\tilde{H}}^{\mathrm{H}}_g = \mathbf{\hat{H}}^\mathrm{H}_g\mathbf{B}_g \in \mathbb{C}^{b_g \times N_g}$ represents the projection of $\mathbf{\hat{H}}_g$ onto the $b_g$--dimensional subspace orthogonal to the $r^* = \sum^{G}_{l\neq g} r_g$ singular vectors associated to the $r_g$ largest singular values of each of the interference groups. In order to distinguish all the $N_g$ users in the group we must have $N_g \leq b_g$, i.e., enough degrees of freedom in the $b_g$--dimensional subspace. Unfortunately, it is not possible to choose $b_g$ and $r_g$ indiscriminately large as one constrains the growth of the other. Specifically, as there exists at most $M$ singular vectors at each group, we have that $N_g \leq b_g \leq M - r^*$. Consequently, a large number of groups leads to less freedom on the choice of both $b_g$ and $r_g$.}

Moreover, $\mathbf{w}_{oc}$, $\mathbf{w}_{ic,g}$ and $\mathbf{w}_{gk} = [\mathbf{W}_{g}]_k$ are the unit norm precoders associated to the instantaneous outer common, inner common and private messages, respectively. We can design 
$\mathbf{W}_g = \xi_{g}\left(\tilde{\mathbf{H}}_g\tilde{\mathbf{H}}_g^\mathrm{H} +  \varepsilon\mathbf{I}_{b_g} \right)^{-1}\tilde{\mathbf{H}}_g$, given a total transmission power $P$, 
as a \gls{rzf} precoder to allow distinguishing between the $N_g$ users within the $g$th group by reducing the interference among the private messages in this group~\cite{adhikary2013_jsdm_large}. The parameter $\xi_g$ is the power normalisation factor which normalizes $||\mathbf{W}_{g}||_2$ to the unit. Likewise, $\varepsilon$ is also a normalisation parameter. Similarly, $\mathbf{w}_{ic,g} = \xi_{ic,g} \sum_{k=1}^{N_g} \mathbf{w}_{gk}$ is the equally weighted \gls{mbf} {built as a linear combination of the private precoders of the $g$th group} where $\xi_{ic,g}$ is a normalisation parameter. Finally, the outer common precoder 
$\mathbf{w}_{oc} = \xi_{oc}\sum_{g=1}^G\sum_{k=1}^{N_g} \mathbf{B}_g\mathbf{\tilde{h}}_{gk}$
is also designed as a weighted \gls{mbf}, but to handle inter-group power leakage where $\xi_{oc}$ is a normalisation paramater. {Notice that it is essential to reduce inter-group interference in order to guarantee communication. Specifically, when group leakage is completely nulled out, there is no need for $\mathbf{w}_{oc}$ and communication happens over $G$ parallel 1-layer \Gls{rs} streams.}

To allocate power among the different messages, we further design two parameters $\alpha, \beta \in (0, 1]$. The first one $\alpha$ represents the fraction of the total power $P$ allocated to the outer common message. And the latter, the  fraction of the remaining power allocated to the inner common message. Combining these, we have $p_{oc} = \alpha P$, $p_{ic, g} = \frac{(1-\alpha)\beta P}{G}$ and $p_{gk} = \frac{(1-\alpha)(1-\beta)P}{N_g}$. In this work we perform a brute force search to find the optimal $\alpha$ and $\beta$ for every channel realisation. 

As mentioned above, at the receiver side, the $k$th user associated to the $g$th group decodes its message in a 2-step successive interference cancellation fashion. In the first step, the user decodes the outer common message ($s_{oc}$) and removes it from the received signal. The group's inner common codeword is then decoded after applying  \gls{sic}. After successfully decoding both common messages, each private message is extracted by considering all other private messages as interference. As a result, the Signal-to-Interference Plus-Noise Ratio (SINR)  to each of these messages is written as 
\begin{align}
\gamma_{gk}^{oc} &= \frac{p_{oc}|\mathbf{h}_{gk}^\mathrm{H}\mathbf{w}_{oc}|^2}{1 + I_{gk}} \\
\gamma_{gk}^{ic} &=
\frac{p_{ic}|\mathbf{{h}}_{gk}^\mathrm{H}\mathbf{w}_{ic, g}|^2}{1 + I_{gk} - p_{ic}|\mathbf{{h}}_{gk}^\mathrm{H}\mathbf{w}_{ic, g}|^2}\\
\gamma_{gk}^{p} &=
\frac{p_{gk}|\mathbf{{h}}_{gk}^\mathrm{H}\mathbf{w}_{gk}|^2}{1 + I_{gk} - \left(p_{ic}|\mathbf{{h}}_{gk}^\mathrm{H}\mathbf{w}_{ic, g}|^2 + p_{gk}|\mathbf{{h}}_{gk}^\mathrm{H}\mathbf{w}_{gk}|^2\right) }
\end{align}
where  $$
I_{gk}~=~\sum_{l = 1}^G p_{ic,l}|\mathbf{h}_{gk}^\mathrm{H}\mathbf{B}_l\mathbf{w}_{ic, l}|^2~+~\sum_{l = 1}^G \sum_{k=1}^{N_g} p_{lk}|\mathbf{h}_{gk}^\mathrm{H}\mathbf{B}_l\mathbf{w}_{lk}|^2
$$
is the combination of all interference leaked from other users and groups. Finally, we can describe the achievable rate as the combination of the smallest achievable outer common rate among all users $R_{oc}=\min\limits_{gk}{\log_2(1+\gamma_{gk}^{oc})}$, the minimal inner common rate per group $R_{ic}=\sum_{g=1}^G\min\limits_k \left(\log_2(1+\gamma_{gk}^{ic})\right)$  and the sum of the rate achievable at all private messages $R_{p}=~\sum_{g=1}^{G}\sum_{k=1}^{N_g} \log_2(1 + \gamma^p_{gk})$. Then the total rate is the sum of these components, i.e, $R = R_{oc} + R_{ic} + R_{p}$. 

\section{User Clustering and Dataset Definition}\label{sec:clustering}
As it becomes evident from the discussion above, and further supported in our results, choosing an appropriate clustering is crucial to take full advantage of two-tier precoding mechanisms, such as \gls{hrs} \cite{adhikary2013_jsdm_large, rs_2rhs}. One can rely on extensive search in order to find the optimal clustering mechanism. However, this is an NP hard task as the number of ways that a set can be partitioned into nonempty sets is given by the Bell number which grows almost exponentially with $N$, i.e., the number of elements in the set. Moreover, in our scenario, many of these partitions lead to vanishing communication rates due to high interference. Therefore, in this work, we rely on (possible suboptimal) clustering options obtained from an agglomerative hierarchical clustering mechanism~\cite{roberto21_globecom}.

\subsection{User Clustering} To devise the clustering mechanism, we define a bottom up approach where the objective is to combine clusters (groups of users in the wireless network) according to their similarity. Initially, each user is associated to a singleton cluster. At each step of the hierarchical clustering algorithm, the pair of users/clusters with highest similarity (according to a criterion discussed later) is then merged. As a result, after each merge we obtain a new clustering option and evaluate the rate achieved considering this new option. This process continues until we have evaluated all levels in the hierarchy. Notice that in this agglomerative mechanism there exist only $N+1$ (total number of users plus one) possible clustering options, one for each level in the hierarchy. These, however, are often relevant clustering options as each cluster only contains elements that are particularly similar to each other.

In~\cite{roberto21_globecom}, we consider the similarity measure between two channel matrices based on how close the principle angles of the subspaces spanned by their column-spaces  are. Specifically, for two clusters of size $N_k$ and $N_j$, we take the projection-Frobenius (PF) similarity 
\begin{equation}
\label{eq:similarity}
    s_{k,j} = \frac{\mathrm{tr}(\hat{\mathbf{P}}_k \hat{\mathbf{P}}_j)}{\min(N_k, N_j)},
\end{equation}
where $\hat{\mathbf{P}}_j$ is the projection matrix given by, 
 \begin{equation}
    \label{eq:preliminary:projection} 
    \hat{\mathbf{P}}_j = \mathbf{\hat{H}}_j(\mathbf{\hat{H}}_j^\mathrm{H} \mathbf{\hat{H}}_j)^{-1}\mathbf{\hat{H}}_j^\mathrm{H}
 \end{equation}
which describes the first $N_j$ left singular vectors of the $k$th group of channels. 
Moreover, to improve clustering results for $N_j \neq N_k$, we follow a statistical analysis of the quantity in (\ref{eq:similarity}) and further define the normalised similarity measure
\begin{equation}
    \label{eq:similarity:normalised}
    \hat{s}_{k,j} = \frac {s_{k,j} - \eta_{k,j}}{\sigma_{k,j}}    
\end{equation}
based on its asymptotic mean $\eta_{k,j}$ and variances $\sigma^2_{k,j}$ defined as in \cite{roberto21_globecom}. However, this normalisation step is only possible for $M > N_j + N_k$, otherwise, we follow the projection-Frobenius similarity described in (\ref{eq:similarity}). 

\subsection{Dataset Definition}
We design the dataset used for this work by devising channel matrices from (\ref{eq:channel:draw}) and clustering them according to the scheme described above. We consider four possible covariance matrices to which channels are randomly associated.  Consequently, for different samples, we might obtain a different number of users associated to a specific covariance matrix. Notice that, this is not a cluster assignment, but merely a way to generate random channels. These covariance matrices are obtained by considering the azimuth angles $\theta_g = -\frac{\pi}{2} + \frac{\pi}{3}(g - 1)$ and the constant angular spread $\Delta_g = \frac{\pi}{6}$. Moreover, we further assume the BS to be equipped with a Uniform Circular Array  antenna.

Concretely, we design 3 different configurations based on the choices of the number of antennas at the BS: 1) $N>M$, 2) $N=M$ and 3) $N < M$. Moreover, we evaluate these configurations for two different system loads, based on the number of users $N \in \{8, 12\}$.  Specifically, we have $M \in \{6, 8, 12\}$ for $N=8$ and $M \in \{6, 12, 16\}$ for $N=12$. As a result, we have 6 different scenarios. For each these we generate $S = 10.000$ random samples, each sample containing both imperfect and perfect CSIT of equal size $N\times M$ and the clustering scheme that maximises the rate based on the hierarchical clustering mechanism. As a result of this randomness, for each scenario we obtain more than $G^* = 200$ possible clustering options, thus, leading to very imbalanced datasets. To diminish this effect, for each scenario, we sub-sample the data such that only relevant classes are left, i.e., we discard classes that achieve less than $25\%$ of the average rate of the scenario and have less than $50$ samples. Moreover, to further balance the data, we crop the maximum number of samples in each class to be at most to $200$. As a result, for each scenario, we still obtain an imbalanced dataset with approximately $G^* = 50$ classes, each containing at least $50$ samples and at most $200$ samples. 

Finally, to compensate for this drop in the number of samples, we further augment the dataset of each configuration by randomly shuffling users that belong to the same cluster. This is a natural extension of this dataset as clustering should be indifferent to the ordering of the users.






\section{Machine learning model and training}

We solve the classification problem presented in the previous section by designing a shallow neural network. We used the Keras library, so we describe the layers with their notation \cite{chollet2015keras}. 
For each scenario, we divide our dataset into training, validation and test sets in a proportion of 80/10/10. During the training procedure, we use the validation set to tune the corresponding hyper-parameters. Our model is defined as a shallow neural network following the parameters from Table \ref{tab:simulation}. The output layer consists of  $G^*$ neurons with a \textit{softmax} activation that correspond to each cluster where $G^*$ is the total number of classes in the scenario.  
The softmax function in the output layer is used to obtain the probability of a user belonging to a specific cluster and it is given by
\begin{equation}
\label{eq:softmax}
\sigma(\mathbf{Z})_g = \frac{e^{z_g}}{\sum_{j=1}^G e^{z_j}}
\end{equation}
where $\mathbf{Z}$ is the input vector from the previous hidden layer, $z_g$ the $g$-th element and the denominator sum is the normalisation factor to ensure the output is into the range of $[0, 1]$. Then, by selecting the maximum, we can obtain the highest probability that users are clustered in a particular way. For the training procedure, we use the Adam optimiser with a learning rate of $10^{-3}$, we train for 50 epochs and use a batch size of 128 samples. For our multi-class classification task, we aim to minimise the categorical cross-entropy loss \cite{goodfellow2016deep}  given by
\begin{equation}
\mathcal{L}(y_g, \hat{y}_g) = - \sum_{g=1}^{G^*} y_g\log\hat{y}_g.
\label{eq:loss}
\end{equation}
where $y_g$ and $\hat{y}_g$ are the groundtruth and NN score for each class. 
This loss is a very good measure of how distinguishable two discrete probability distributions are from each other. In this context, the vector $\mathbf{\hat{y}} = [\hat{y}_1, \dots, \hat{y}_g] \in \mathbb{R}^{G^*}$ has entries which represent the probability that users are clustered in a specific manner and the sum of all entries is one. The accuracy of a model is often defined in terms of the entry with highest probability, this is often, called \textit{top-1} accuracy. In our scenario, there exist several options which achieve the (close to) maximum rate. Therefore, it is also interesting to analyse the \textit{top-k} accuracy of our model, i.e., if the desired clustering option is among the $k$ most probable outputs.  


Finally, we emphasise that we are applying a shallow neural network which contains only a small number of learnable parameters. This is designed as a consequence of our devised dataset. Recall that we have specifically defined it to be imbalanced and with a small number of samples to each class ($[50,200]$). Nonetheless, as we present below, this network is capable of learning the relationship between the different channel matrices and directly output the desired clustering option that maximises transmission using \Gls{hrs}. 

\section{Performance Analysis}

\begin{table}
\centering
\caption{Parameters of the Simulations}
\begin{tabular}{|c|c|}
\hline
\textbf{Simulation Parameter}         & \textbf{Simulation Value}            \\ \hline
Antenna Configuration                & Uniform Circular Array         \\ \hline
Angular Spread ($\Delta_g$)              & ${\pi}/{6}$ \\ \hline
Number of Unique Distributions                & 4 \\ \hline
Channel Quality ($\tau^2)$                & 0.4 \\ \hline
Dominant Eigenvectors ($b_g=r_g$)                & $ \lfloor M/G \rfloor$ \\ \hline
Channel Quality ($r_g$)                & 0.4 \\ \hline
Number Shuffling              & 10 \\ \hline
Number of Neurons in NN          &  $\{256, 128\}$                      \\ \hline
NN Learning Rate & $10^{-3}$                         \\ \hline
NN Training Epochs & $50$                         \\ \hline
NN Training Batch Size & $128$                         \\ \hline
NN Input Layer Activation Function & ReLu Function                         \\ \hline
NN Hidden Layer Activation Function & ReLu Function                         \\ \hline
NN Output Layer Activation Function & Softmax Function                         \\ \hline
NN Loss Function & Categorical Cross-entropy Loss                       \\ \hline
\end{tabular}
\label{tab:simulation}
\vspace{-1\baselineskip}
\end{table}
In this section, we evaluate the performance of the presented \gls{nn} method in comparison to \gls{rs} under different scenario configurations. These numerical simulations are carried out in a MATLAB environment. The necessary configuration parameters are defined in Table \ref{tab:simulation}.



In order to validate the learning of the \gls{nn}, we compare the rate achieved using the \gls{nn} predicted classes and different \gls{rs} clustering options. To perform a complete evaluation, we determine the rate achieved by the following solutions,
\begin{itemize}
    \item Hierarchical Clustering - Hierarchical Rate Splitting (HC): The users are clustered according to the clustering mechanism defined in Sec. \ref{sec:clustering}, the group with higher communication performance is selected;
    \item Neural Network - Hierarchical Rate Splitting (NN): Proposed \gls{nn} based clustering;
    \item Universal Cluster (UNI): All users are clustered into one single cluster;
    \item Singleton Cluster (SING): Each cluster contains only single user.
\end{itemize}

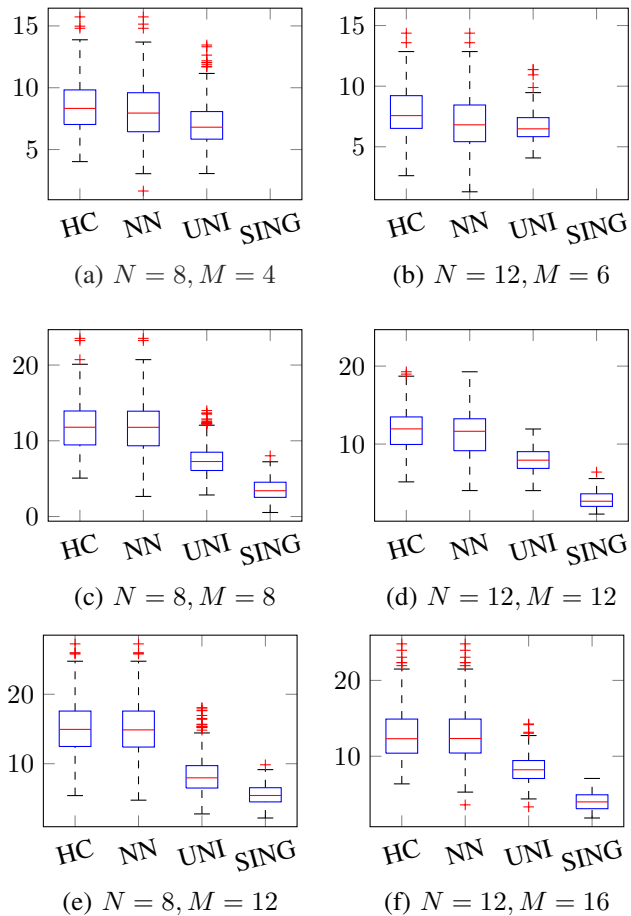
\begin{figure}[!ht]
     \centering
     \begin{subfigure}[h]{0.23\textwidth}
         \centering
%
%
\begin{tikzpicture}

\begin{axis}[%
width=0.8\linewidth,
height=1.0in,
at={(0in,1.5in)},
scale only axis,
unbounded coords=jump,
xlabel style={font=\color{white!15!black}},
xmin=0.5,
xmax=4.5,
xtick={1,2,3,4},
xticklabels={{HC},{NN},{UNI},{SING}},
xticklabel style={rotate=15},
ymin=0.94169938584686,
ymax=16.43524796451,
axis background/.style={fill=white},
title style={font=\bfseries},
xlabel={(a) $N=8,M=4$},
]
\addplot [color=black, dashed, forget plot]
  table[row sep=crcr]{%
1	9.82412890174763\\
1	13.8789892934799\\
};
\addplot [color=black, dashed, forget plot]
  table[row sep=crcr]{%
2	9.59726525191918\\
2	13.6950222145863\\
};
\addplot [color=black, dashed, forget plot]
  table[row sep=crcr]{%
3	8.07315767428321\\
3	11.1564671457031\\
};
\addplot [color=black, dashed, forget plot]
  table[row sep=crcr]{%
4	-inf\\
4	-inf\\
};
\addplot [color=black, dashed, forget plot]
  table[row sep=crcr]{%
1	4.02526549607622\\
1	7.0288446845198\\
};
\addplot [color=black, dashed, forget plot]
  table[row sep=crcr]{%
2	3.04836947425197\\
2	6.44156082873789\\
};
\addplot [color=black, dashed, forget plot]
  table[row sep=crcr]{%
3	3.05941386823017\\
3	5.84076844528711\\
};
\addplot [color=black, dashed, forget plot]
  table[row sep=crcr]{%
4	-inf\\
4	-inf\\
};
\addplot [color=black, forget plot]
  table[row sep=crcr]{%
0.875	13.8789892934799\\
1.125	13.8789892934799\\
};
\addplot [color=black, forget plot]
  table[row sep=crcr]{%
1.875	13.6950222145863\\
2.125	13.6950222145863\\
};
\addplot [color=black, forget plot]
  table[row sep=crcr]{%
2.875	11.1564671457031\\
3.125	11.1564671457031\\
};
\addplot [color=black, forget plot]
  table[row sep=crcr]{%
3.875	-inf\\
4.125	-inf\\
};
\addplot [color=black, forget plot]
  table[row sep=crcr]{%
0.875	4.02526549607622\\
1.125	4.02526549607622\\
};
\addplot [color=black, forget plot]
  table[row sep=crcr]{%
1.875	3.04836947425197\\
2.125	3.04836947425197\\
};
\addplot [color=black, forget plot]
  table[row sep=crcr]{%
2.875	3.05941386823017\\
3.125	3.05941386823017\\
};
\addplot [color=black, forget plot]
  table[row sep=crcr]{%
3.875	-inf\\
4.125	-inf\\
};
\addplot [color=blue, forget plot]
  table[row sep=crcr]{%
0.75	7.0288446845198\\
0.75	9.82412890174763\\
1.25	9.82412890174763\\
1.25	7.0288446845198\\
0.75	7.0288446845198\\
};
\addplot [color=blue, forget plot]
  table[row sep=crcr]{%
1.75	6.44156082873789\\
1.75	9.59726525191918\\
2.25	9.59726525191918\\
2.25	6.44156082873789\\
1.75	6.44156082873789\\
};
\addplot [color=blue, forget plot]
  table[row sep=crcr]{%
2.75	5.84076844528711\\
2.75	8.07315767428321\\
3.25	8.07315767428321\\
3.25	5.84076844528711\\
2.75	5.84076844528711\\
};
\addplot [color=blue, forget plot]
  table[row sep=crcr]{%
3.75	-inf\\
3.75	-inf\\
4.25	-inf\\
4.25	-inf\\
3.75	-inf\\
};
\addplot [color=red, forget plot]
  table[row sep=crcr]{%
0.75	8.32831967667055\\
1.25	8.32831967667055\\
};
\addplot [color=red, forget plot]
  table[row sep=crcr]{%
1.75	7.95012878293102\\
2.25	7.95012878293102\\
};
\addplot [color=red, forget plot]
  table[row sep=crcr]{%
2.75	6.80864488536246\\
3.25	6.80864488536246\\
};
\addplot [color=red, forget plot]
  table[row sep=crcr]{%
3.75	-inf\\
4.25	-inf\\
};
\addplot [color=black, only marks, mark=+, mark options={solid, draw=red}, forget plot]
  table[row sep=crcr]{%
1	14.8097613857525\\
1	14.9804388901998\\
1	15.7309957563889\\
};
\addplot [color=black, only marks, mark=+, mark options={solid, draw=red}, forget plot]
  table[row sep=crcr]{%
2	1.64595159396791\\
2	14.8097613857525\\
2	15.1464475176126\\
2	15.7309957563889\\
};
\addplot [color=black, only marks, mark=+, mark options={solid, draw=red}, forget plot]
  table[row sep=crcr]{%
3	11.6885867468306\\
3	11.7152424765091\\
3	11.8619510804515\\
3	12.0287015711463\\
3	12.0374276909113\\
3	12.1793943390743\\
3	12.6385685702889\\
3	13.3156304489042\\
3	13.4566723941054\\
};
\addplot [color=black, only marks, mark=+, mark options={solid, draw=red}, forget plot]
  table[row sep=crcr]{%
nan	nan\\
};
\end{axis}
\end{tikzpicture}%
     \end{subfigure}
     \begin{subfigure}[h]{0.23\textwidth}
         \centering
%
%
\begin{tikzpicture}

\begin{axis}[%
width=0.8\linewidth,
height=1.0in,
at={(0in,1.5in)},
scale only axis,
unbounded coords=jump,
xmin=0.5,
xmax=4.5,
xtick={1,2,3,4},
xticklabels={{HC},{NN},{UNI},{SING}},
xticklabel style={rotate=15},
ymin=0.630648258821232,
ymax=16.43524796451,
axis background/.style={fill=white},
title style={font=\bfseries},
xlabel={(b) $N=12,M=6$},
]
\addplot [color=black, dashed, forget plot]
  table[row sep=crcr]{%
1	9.21714157574479\\
1	12.8551257318123\\
};
\addplot [color=black, dashed, forget plot]
  table[row sep=crcr]{%
2	8.44680527305263\\
2	12.8551257318123\\
};
\addplot [color=black, dashed, forget plot]
  table[row sep=crcr]{%
3	7.40454582026483\\
3	9.46841946781415\\
};
\addplot [color=black, dashed, forget plot]
  table[row sep=crcr]{%
4	-inf\\
4	-inf\\
};
\addplot [color=black, dashed, forget plot]
  table[row sep=crcr]{%
1	2.61517595264601\\
1	6.51655670103332\\
};
\addplot [color=black, dashed, forget plot]
  table[row sep=crcr]{%
2	1.2851632853337\\
2	5.42108140801899\\
};
\addplot [color=black, dashed, forget plot]
  table[row sep=crcr]{%
3	4.07756433164607\\
3	5.83394458479291\\
};
\addplot [color=black, dashed, forget plot]
  table[row sep=crcr]{%
4	-inf\\
4	-inf\\
};
\addplot [color=black, forget plot]
  table[row sep=crcr]{%
0.875	12.8551257318123\\
1.125	12.8551257318123\\
};
\addplot [color=black, forget plot]
  table[row sep=crcr]{%
1.875	12.8551257318123\\
2.125	12.8551257318123\\
};
\addplot [color=black, forget plot]
  table[row sep=crcr]{%
2.875	9.46841946781415\\
3.125	9.46841946781415\\
};
\addplot [color=black, forget plot]
  table[row sep=crcr]{%
3.875	-inf\\
4.125	-inf\\
};
\addplot [color=black, forget plot]
  table[row sep=crcr]{%
0.875	2.61517595264601\\
1.125	2.61517595264601\\
};
\addplot [color=black, forget plot]
  table[row sep=crcr]{%
1.875	1.2851632853337\\
2.125	1.2851632853337\\
};
\addplot [color=black, forget plot]
  table[row sep=crcr]{%
2.875	4.07756433164607\\
3.125	4.07756433164607\\
};
\addplot [color=black, forget plot]
  table[row sep=crcr]{%
3.875	-inf\\
4.125	-inf\\
};
\addplot [color=blue, forget plot]
  table[row sep=crcr]{%
0.75	6.51655670103332\\
0.75	9.21714157574479\\
1.25	9.21714157574479\\
1.25	6.51655670103332\\
0.75	6.51655670103332\\
};
\addplot [color=blue, forget plot]
  table[row sep=crcr]{%
1.75	5.42108140801899\\
1.75	8.44680527305263\\
2.25	8.44680527305263\\
2.25	5.42108140801899\\
1.75	5.42108140801899\\
};
\addplot [color=blue, forget plot]
  table[row sep=crcr]{%
2.75	5.83394458479291\\
2.75	7.40454582026483\\
3.25	7.40454582026483\\
3.25	5.83394458479291\\
2.75	5.83394458479291\\
};
\addplot [color=blue, forget plot]
  table[row sep=crcr]{%
3.75	-inf\\
3.75	-inf\\
4.25	-inf\\
4.25	-inf\\
3.75	-inf\\
};
\addplot [color=red, forget plot]
  table[row sep=crcr]{%
0.75	7.57103421926809\\
1.25	7.57103421926809\\
};
\addplot [color=red, forget plot]
  table[row sep=crcr]{%
1.75	6.81081648032711\\
2.25	6.81081648032711\\
};
\addplot [color=red, forget plot]
  table[row sep=crcr]{%
2.75	6.48019675648186\\
3.25	6.48019675648186\\
};
\addplot [color=red, forget plot]
  table[row sep=crcr]{%
3.75	-inf\\
4.25	-inf\\
};
\addplot [color=black, only marks, mark=+, mark options={solid, draw=red}, forget plot]
  table[row sep=crcr]{%
1	13.5863568112946\\
1	14.375463815583\\
};
\addplot [color=black, only marks, mark=+, mark options={solid, draw=red}, forget plot]
  table[row sep=crcr]{%
2	13.5863568112946\\
2	14.375463815583\\
};
\addplot [color=black, only marks, mark=+, mark options={solid, draw=red}, forget plot]
  table[row sep=crcr]{%
3	9.88967570109632\\
3	10.9355535016396\\
3	11.3648388319619\\
3	11.3772246327567\\
};
\addplot [color=black, only marks, mark=+, mark options={solid, draw=red}, forget plot]
  table[row sep=crcr]{%
nan	nan\\
};
\end{axis}
\end{tikzpicture}%
     \end{subfigure}
     \vspace{-0.5\baselineskip}
     \begin{subfigure}[t]{0.23\textwidth}
         \centering
%
%
\begin{tikzpicture}

\begin{axis}[%
width=0.8\linewidth,
height=1.0in,
at={(0in,1.5in)},
scale only axis,
xmin=0.5,
xmax=4.5,
xtick={1,2,3,4},
xticklabels={{HC},{NN},{UNI},{SING}},
xticklabel style={rotate=15},
ymin=-0.615342775844212,
ymax=24.6751521308473,
axis background/.style={fill=white},
title style={font=\bfseries},
xlabel={(c) $N=8, M=8$},
]
\addplot [color=black, dashed, forget plot]
  table[row sep=crcr]{%
1	13.9366493864716\\
1	20.1031090550873\\
};
\addplot [color=black, dashed, forget plot]
  table[row sep=crcr]{%
2	13.9085851797276\\
2	20.7158603231912\\
};
\addplot [color=black, dashed, forget plot]
  table[row sep=crcr]{%
3	8.49250722875024\\
3	12.0579400377951\\
};
\addplot [color=black, dashed, forget plot]
  table[row sep=crcr]{%
4	4.53056430521348\\
4	7.232335830705\\
};
\addplot [color=black, dashed, forget plot]
  table[row sep=crcr]{%
1	5.07351907446347\\
1	9.45639372072728\\
};
\addplot [color=black, dashed, forget plot]
  table[row sep=crcr]{%
2	2.6535796538968\\
2	9.34278747009685\\
};
\addplot [color=black, dashed, forget plot]
  table[row sep=crcr]{%
3	2.84780926451592\\
3	6.07881053786764\\
};
\addplot [color=black, dashed, forget plot]
  table[row sep=crcr]{%
4	0.534225174459947\\
4	2.53058925997957\\
};
\addplot [color=black, forget plot]
  table[row sep=crcr]{%
0.875	20.1031090550873\\
1.125	20.1031090550873\\
};
\addplot [color=black, forget plot]
  table[row sep=crcr]{%
1.875	20.7158603231912\\
2.125	20.7158603231912\\
};
\addplot [color=black, forget plot]
  table[row sep=crcr]{%
2.875	12.0579400377951\\
3.125	12.0579400377951\\
};
\addplot [color=black, forget plot]
  table[row sep=crcr]{%
3.875	7.232335830705\\
4.125	7.232335830705\\
};
\addplot [color=black, forget plot]
  table[row sep=crcr]{%
0.875	5.07351907446347\\
1.125	5.07351907446347\\
};
\addplot [color=black, forget plot]
  table[row sep=crcr]{%
1.875	2.6535796538968\\
2.125	2.6535796538968\\
};
\addplot [color=black, forget plot]
  table[row sep=crcr]{%
2.875	2.84780926451592\\
3.125	2.84780926451592\\
};
\addplot [color=black, forget plot]
  table[row sep=crcr]{%
3.875	0.534225174459947\\
4.125	0.534225174459947\\
};
\addplot [color=blue, forget plot]
  table[row sep=crcr]{%
0.75	9.45639372072728\\
0.75	13.9366493864716\\
1.25	13.9366493864716\\
1.25	9.45639372072728\\
0.75	9.45639372072728\\
};
\addplot [color=blue, forget plot]
  table[row sep=crcr]{%
1.75	9.34278747009685\\
1.75	13.9085851797276\\
2.25	13.9085851797276\\
2.25	9.34278747009685\\
1.75	9.34278747009685\\
};
\addplot [color=blue, forget plot]
  table[row sep=crcr]{%
2.75	6.07881053786764\\
2.75	8.49250722875024\\
3.25	8.49250722875024\\
3.25	6.07881053786764\\
2.75	6.07881053786764\\
};
\addplot [color=blue, forget plot]
  table[row sep=crcr]{%
3.75	2.53058925997957\\
3.75	4.53056430521348\\
4.25	4.53056430521348\\
4.25	2.53058925997957\\
3.75	2.53058925997957\\
};
\addplot [color=red, forget plot]
  table[row sep=crcr]{%
0.75	11.790199306073\\
1.25	11.790199306073\\
};
\addplot [color=red, forget plot]
  table[row sep=crcr]{%
1.75	11.7806003894234\\
2.25	11.7806003894234\\
};
\addplot [color=red, forget plot]
  table[row sep=crcr]{%
2.75	7.26420965652891\\
3.25	7.26420965652891\\
};
\addplot [color=red, forget plot]
  table[row sep=crcr]{%
3.75	3.40689760151722\\
4.25	3.40689760151722\\
};
\addplot [color=black, only marks, mark=+, mark options={solid, draw=red}, forget plot]
  table[row sep=crcr]{%
1	20.7158603231912\\
1	23.231302049531\\
1	23.5255841805431\\
};
\addplot [color=black, only marks, mark=+, mark options={solid, draw=red}, forget plot]
  table[row sep=crcr]{%
2	23.231302049531\\
2	23.5255841805431\\
};
\addplot [color=black, only marks, mark=+, mark options={solid, draw=red}, forget plot]
  table[row sep=crcr]{%
3	12.1945481728484\\
3	12.2274247372482\\
3	12.2646126752421\\
3	12.3365944488233\\
3	12.3730250593822\\
3	12.393531561095\\
3	12.5360403441556\\
3	12.7528106210332\\
3	12.8813693034901\\
3	13.5157379686191\\
3	13.6436475140016\\
3	13.6652084101394\\
3	13.9913615660886\\
};
\addplot [color=black, only marks, mark=+, mark options={solid, draw=red}, forget plot]
  table[row sep=crcr]{%
4	8.02214132465536\\
};
\end{axis}
\end{tikzpicture}%
     \end{subfigure}
     \begin{subfigure}[t]{0.23\textwidth}
         \centering
%
%
\begin{tikzpicture}

\begin{axis}[%
width=0.8\linewidth,
height=1.0in,
at={(0in,1.5in)},
scale only axis,
unbounded coords=jump,
xmin=0.5,
xmax=4.5,
xtick={1,2,3,4},
xticklabels={{HC},{NN},{UNI},{SING}},
xticklabel style={rotate=15},
ymin=0.105321393640342,
ymax=24.6751521308473,
axis background/.style={fill=white},
title style={font=\bfseries},
xlabel={(d) $N=12, M=12$},
]
\addplot [color=black, dashed, forget plot]
  table[row sep=crcr]{%
1	13.4795022341395\\
1	18.7056579992161\\
};
\addplot [color=black, dashed, forget plot]
  table[row sep=crcr]{%
2	13.2420620803227\\
2	19.2736740549296\\
};
\addplot [color=black, dashed, forget plot]
  table[row sep=crcr]{%
3	9.02742747053766\\
3	11.9318076717776\\
};
\addplot [color=black, dashed, forget plot]
  table[row sep=crcr]{%
4	3.6181887445671\\
4	5.57488491353344\\
};
\addplot [color=black, dashed, forget plot]
  table[row sep=crcr]{%
1	5.15008574325767\\
1	9.94775412122578\\
};
\addplot [color=black, dashed, forget plot]
  table[row sep=crcr]{%
2	4.03319741847725\\
2	9.14818270321123\\
};
\addplot [color=black, dashed, forget plot]
  table[row sep=crcr]{%
3	4.02715323761661\\
3	6.87613795422178\\
};
\addplot [color=black, dashed, forget plot]
  table[row sep=crcr]{%
4	1.01810009179697\\
4	2.00613971239212\\
};
\addplot [color=black, forget plot]
  table[row sep=crcr]{%
0.875	18.7056579992161\\
1.125	18.7056579992161\\
};
\addplot [color=black, forget plot]
  table[row sep=crcr]{%
1.875	19.2736740549296\\
2.125	19.2736740549296\\
};
\addplot [color=black, forget plot]
  table[row sep=crcr]{%
2.875	11.9318076717776\\
3.125	11.9318076717776\\
};
\addplot [color=black, forget plot]
  table[row sep=crcr]{%
3.875	5.57488491353344\\
4.125	5.57488491353344\\
};
\addplot [color=black, forget plot]
  table[row sep=crcr]{%
0.875	5.15008574325767\\
1.125	5.15008574325767\\
};
\addplot [color=black, forget plot]
  table[row sep=crcr]{%
1.875	4.03319741847725\\
2.125	4.03319741847725\\
};
\addplot [color=black, forget plot]
  table[row sep=crcr]{%
2.875	4.02715323761661\\
3.125	4.02715323761661\\
};
\addplot [color=black, forget plot]
  table[row sep=crcr]{%
3.875	1.01810009179697\\
4.125	1.01810009179697\\
};
\addplot [color=blue, forget plot]
  table[row sep=crcr]{%
0.75	9.94775412122578\\
0.75	13.4795022341395\\
1.25	13.4795022341395\\
1.25	9.94775412122578\\
0.75	9.94775412122578\\
};
\addplot [color=blue, forget plot]
  table[row sep=crcr]{%
1.75	9.14818270321123\\
1.75	13.2420620803227\\
2.25	13.2420620803227\\
2.25	9.14818270321123\\
1.75	9.14818270321123\\
};
\addplot [color=blue, forget plot]
  table[row sep=crcr]{%
2.75	6.87613795422178\\
2.75	9.02742747053766\\
3.25	9.02742747053766\\
3.25	6.87613795422178\\
2.75	6.87613795422178\\
};
\addplot [color=blue, forget plot]
  table[row sep=crcr]{%
3.75	2.00613971239212\\
3.75	3.6181887445671\\
4.25	3.6181887445671\\
4.25	2.00613971239212\\
3.75	2.00613971239212\\
};
\addplot [color=red, forget plot]
  table[row sep=crcr]{%
0.75	11.9458723242437\\
1.25	11.9458723242437\\
};
\addplot [color=red, forget plot]
  table[row sep=crcr]{%
1.75	11.6386642027206\\
2.25	11.6386642027206\\
};
\addplot [color=red, forget plot]
  table[row sep=crcr]{%
2.75	7.93303333088752\\
3.25	7.93303333088752\\
};
\addplot [color=red, forget plot]
  table[row sep=crcr]{%
3.75	2.66491196782328\\
4.25	2.66491196782328\\
};
\addplot [color=black, only marks, mark=+, mark options={solid, draw=red}, forget plot]
  table[row sep=crcr]{%
1	18.9732108604706\\
1	19.2736740549296\\
};
\addplot [color=black, only marks, mark=+, mark options={solid, draw=red}, forget plot]
  table[row sep=crcr]{%
nan	nan\\
};
\addplot [color=black, only marks, mark=+, mark options={solid, draw=red}, forget plot]
  table[row sep=crcr]{%
nan	nan\\
};
\addplot [color=black, only marks, mark=+, mark options={solid, draw=red}, forget plot]
  table[row sep=crcr]{%
4	6.40659239929579\\
};
\end{axis}

\end{tikzpicture}%
     \end{subfigure}
     \vspace{-0.5\baselineskip}
    \begin{subfigure}[t]{0.23\textwidth}
         \centering
%
%
\begin{tikzpicture}

\begin{axis}[%
width=0.8\linewidth,
height=1.0in,
at={(0in,1.5in)},
scale only axis,
xmin=0.5,
xmax=4.5,
xtick={1,2,3,4},
xticklabels={{HC},{NN},{UNI},{SING}},
xticklabel style={rotate=15},
ymin=0.931480904796599,
ymax=28.5062136083999,
axis background/.style={fill=white},
title style={font=\bfseries},
xlabel={(e) $N=8, M=12$},
]
\addplot [color=black, dashed, forget plot]
  table[row sep=crcr]{%
1	17.5795584435468\\
1	24.7529308546466\\
};
\addplot [color=black, dashed, forget plot]
  table[row sep=crcr]{%
2	17.5795584435468\\
2	24.7529308546466\\
};
\addplot [color=black, dashed, forget plot]
  table[row sep=crcr]{%
3	9.72520023460109\\
3	14.429443133594\\
};
\addplot [color=black, dashed, forget plot]
  table[row sep=crcr]{%
4	6.54539186355954\\
4	9.15138245731808\\
};
\addplot [color=black, dashed, forget plot]
  table[row sep=crcr]{%
1	5.41044896273751\\
1	12.4817532351226\\
};
\addplot [color=black, dashed, forget plot]
  table[row sep=crcr]{%
2	4.7520891277269\\
2	12.3994932911933\\
};
\addplot [color=black, dashed, forget plot]
  table[row sep=crcr]{%
3	2.77217065973193\\
3	6.50013560193885\\
};
\addplot [color=black, dashed, forget plot]
  table[row sep=crcr]{%
4	2.18487784586948\\
4	4.50444233735167\\
};
\addplot [color=black, forget plot]
  table[row sep=crcr]{%
0.875	24.7529308546466\\
1.125	24.7529308546466\\
};
\addplot [color=black, forget plot]
  table[row sep=crcr]{%
1.875	24.7529308546466\\
2.125	24.7529308546466\\
};
\addplot [color=black, forget plot]
  table[row sep=crcr]{%
2.875	14.429443133594\\
3.125	14.429443133594\\
};
\addplot [color=black, forget plot]
  table[row sep=crcr]{%
3.875	9.15138245731808\\
4.125	9.15138245731808\\
};
\addplot [color=black, forget plot]
  table[row sep=crcr]{%
0.875	5.41044896273751\\
1.125	5.41044896273751\\
};
\addplot [color=black, forget plot]
  table[row sep=crcr]{%
1.875	4.7520891277269\\
2.125	4.7520891277269\\
};
\addplot [color=black, forget plot]
  table[row sep=crcr]{%
2.875	2.77217065973193\\
3.125	2.77217065973193\\
};
\addplot [color=black, forget plot]
  table[row sep=crcr]{%
3.875	2.18487784586948\\
4.125	2.18487784586948\\
};
\addplot [color=blue, forget plot]
  table[row sep=crcr]{%
0.75	12.4817532351226\\
0.75	17.5795584435468\\
1.25	17.5795584435468\\
1.25	12.4817532351226\\
0.75	12.4817532351226\\
};
\addplot [color=blue, forget plot]
  table[row sep=crcr]{%
1.75	12.3994932911933\\
1.75	17.5795584435468\\
2.25	17.5795584435468\\
2.25	12.3994932911933\\
1.75	12.3994932911933\\
};
\addplot [color=blue, forget plot]
  table[row sep=crcr]{%
2.75	6.50013560193885\\
2.75	9.72520023460109\\
3.25	9.72520023460109\\
3.25	6.50013560193885\\
2.75	6.50013560193885\\
};
\addplot [color=blue, forget plot]
  table[row sep=crcr]{%
3.75	4.50444233735167\\
3.75	6.54539186355954\\
4.25	6.54539186355954\\
4.25	4.50444233735167\\
3.75	4.50444233735167\\
};
\addplot [color=red, forget plot]
  table[row sep=crcr]{%
0.75	14.9423251826336\\
1.25	14.9423251826336\\
};
\addplot [color=red, forget plot]
  table[row sep=crcr]{%
1.75	14.8733799450787\\
2.25	14.8733799450787\\
};
\addplot [color=red, forget plot]
  table[row sep=crcr]{%
2.75	7.96839957407168\\
3.25	7.96839957407168\\
};
\addplot [color=red, forget plot]
  table[row sep=crcr]{%
3.75	5.42501586520637\\
4.25	5.42501586520637\\
};
\addplot [color=black, only marks, mark=+, mark options={solid, draw=red}, forget plot]
  table[row sep=crcr]{%
1	25.7945899390648\\
1	25.903941439068\\
1	25.9750050574615\\
1	27.2528166673271\\
};
\addplot [color=black, only marks, mark=+, mark options={solid, draw=red}, forget plot]
  table[row sep=crcr]{%
2	25.7945899390648\\
2	25.903941439068\\
2	25.9750050574615\\
2	27.2528166673271\\
};
\addplot [color=black, only marks, mark=+, mark options={solid, draw=red}, forget plot]
  table[row sep=crcr]{%
3	14.8320977293615\\
3	15.1947597930686\\
3	15.2208908062875\\
3	15.2808709715223\\
3	15.3484991140627\\
3	15.3981254578909\\
3	15.6388891446493\\
3	15.6800803128254\\
3	16.3959947269028\\
3	16.511079937502\\
3	16.9447349933629\\
3	17.6331650620779\\
3	17.6949226378309\\
3	18.0200153169294\\
3	18.0612944661336\\
};
\addplot [color=black, only marks, mark=+, mark options={solid, draw=red}, forget plot]
  table[row sep=crcr]{%
4	9.84692496831532\\
};
\end{axis}
\end{tikzpicture}%
     \end{subfigure}
     \begin{subfigure}[t]{0.23\textwidth}
         \centering
%
%
\begin{tikzpicture}

\begin{axis}[%
width=0.8\linewidth,
height=1.0in,
at={(0in,1.5in)},
scale only axis,
unbounded coords=jump,
xmin=0.5,
xmax=4.5,
xtick={1,2,3,4},
xticklabels={{HC},{NN},{UNI},{SING}},
xticklabel style={rotate=15},
ymin=0.710851036947486,
ymax=25.9596957350296,
axis background/.style={fill=white},
title style={font=\bfseries},
xlabel={(f) $N=12, M=16$},
]
\addplot [color=black, dashed, forget plot]
  table[row sep=crcr]{%
1	14.8987157925543\\
1	21.4928462831547\\
};
\addplot [color=black, dashed, forget plot]
  table[row sep=crcr]{%
2	14.8987157925543\\
2	21.4928462831547\\
};
\addplot [color=black, dashed, forget plot]
  table[row sep=crcr]{%
3	9.43338780998462\\
3	12.726816142831\\
};
\addplot [color=black, dashed, forget plot]
  table[row sep=crcr]{%
4	4.91976282265388\\
4	7.07226200280526\\
};
\addplot [color=black, dashed, forget plot]
  table[row sep=crcr]{%
1	6.35860503346581\\
1	10.402964274849\\
};
\addplot [color=black, dashed, forget plot]
  table[row sep=crcr]{%
2	5.27248009670467\\
2	10.4238041978446\\
};
\addplot [color=black, dashed, forget plot]
  table[row sep=crcr]{%
3	4.37334630412094\\
3	7.06837767135583\\
};
\addplot [color=black, dashed, forget plot]
  table[row sep=crcr]{%
4	1.85852579595122\\
4	3.0854805140385\\
};
\addplot [color=black, forget plot]
  table[row sep=crcr]{%
0.875	21.4928462831547\\
1.125	21.4928462831547\\
};
\addplot [color=black, forget plot]
  table[row sep=crcr]{%
1.875	21.4928462831547\\
2.125	21.4928462831547\\
};
\addplot [color=black, forget plot]
  table[row sep=crcr]{%
2.875	12.726816142831\\
3.125	12.726816142831\\
};
\addplot [color=black, forget plot]
  table[row sep=crcr]{%
3.875	7.07226200280526\\
4.125	7.07226200280526\\
};
\addplot [color=black, forget plot]
  table[row sep=crcr]{%
0.875	6.35860503346581\\
1.125	6.35860503346581\\
};
\addplot [color=black, forget plot]
  table[row sep=crcr]{%
1.875	5.27248009670467\\
2.125	5.27248009670467\\
};
\addplot [color=black, forget plot]
  table[row sep=crcr]{%
2.875	4.37334630412094\\
3.125	4.37334630412094\\
};
\addplot [color=black, forget plot]
  table[row sep=crcr]{%
3.875	1.85852579595122\\
4.125	1.85852579595122\\
};
\addplot [color=blue, forget plot]
  table[row sep=crcr]{%
0.75	10.402964274849\\
0.75	14.8987157925543\\
1.25	14.8987157925543\\
1.25	10.402964274849\\
0.75	10.402964274849\\
};
\addplot [color=blue, forget plot]
  table[row sep=crcr]{%
1.75	10.4238041978446\\
1.75	14.8987157925543\\
2.25	14.8987157925543\\
2.25	10.4238041978446\\
1.75	10.4238041978446\\
};
\addplot [color=blue, forget plot]
  table[row sep=crcr]{%
2.75	7.06837767135583\\
2.75	9.43338780998462\\
3.25	9.43338780998462\\
3.25	7.06837767135583\\
2.75	7.06837767135583\\
};
\addplot [color=blue, forget plot]
  table[row sep=crcr]{%
3.75	3.0854805140385\\
3.75	4.91976282265388\\
4.25	4.91976282265388\\
4.25	3.0854805140385\\
3.75	3.0854805140385\\
};
\addplot [color=red, forget plot]
  table[row sep=crcr]{%
0.75	12.3111323342272\\
1.25	12.3111323342272\\
};
\addplot [color=red, forget plot]
  table[row sep=crcr]{%
1.75	12.3374520780579\\
2.25	12.3374520780579\\
};
\addplot [color=red, forget plot]
  table[row sep=crcr]{%
2.75	8.21664878281523\\
3.25	8.21664878281523\\
};
\addplot [color=red, forget plot]
  table[row sep=crcr]{%
3.75	3.9820625393566\\
4.25	3.9820625393566\\
};
\addplot [color=black, only marks, mark=+, mark options={solid, draw=red}, forget plot]
  table[row sep=crcr]{%
1	21.9443971876473\\
1	22.3404074248652\\
1	22.3483811674139\\
1	23.0515489042612\\
1	23.9691613359589\\
1	24.8120209760259\\
};
\addplot [color=black, only marks, mark=+, mark options={solid, draw=red}, forget plot]
  table[row sep=crcr]{%
2	3.59055065210665\\
2	21.9443971876473\\
2	22.3404074248652\\
2	22.3483811674139\\
2	23.0515489042612\\
2	23.9691613359589\\
2	24.8120209760259\\
};
\addplot [color=black, only marks, mark=+, mark options={solid, draw=red}, forget plot]
  table[row sep=crcr]{%
3	3.32393528676412\\
3	13.0279251616269\\
3	13.1587037204489\\
3	14.1644527171919\\
3	14.2551150142185\\
};
\end{axis}
\end{tikzpicture}%
     \end{subfigure}
     \vspace{-0.5\baselineskip}
     \caption{Spectral efficiency (bps/Hz) achieved for clustering mechanisms using HRS. }
    \label{fig:boxplots}
\end{figure}
As mentioned above, we consider three scenarios to evaluate the clustering solutions 
\begin{enumerate*}
    \item $M < N$, \item $M = N$ and \item $M > N$.
\end{enumerate*} 
Hence, for $N=8$, we determine the rate achieved for $M\in\{4,8,12\}$ and for $N=12$, we determine the rate achieved for $M\in\{6,12,16\}$. Then, we compare the different clustering techniques mentioned before based on the rate achieved. \figurename~\ref{fig:boxplots} shows the rate achieved for all four clustering techniques for the different values of $M$ and $N$. {Each box plot shows the rate obtained for different realisations of the channel. The median rate is presented by a horizontal line through box and the top and bottom of the box are the $75$th and $25$th percentile rate (i.e. rate achieved by $75\%$ and $25\%$ of the scenarios). Lastly, the extremities of the boxplot refer to the $1\%$ and $99\%$ and the red plus indicators in the boxplot denote the outlier rate values.}   Notice that the rate achieved by HC-HRS and NN-HRS is approximately similar while both clustering techniques outperform UNI and SING. This is due to the fact that with a noisy channel, it is really difficult to generate accurate precoders that can maximise the rate achieved and minimise the inter-group and intra-group interferences. {Additionally, the NN-HRS only receives the instantaneous noisy channel as an input and determines its clustering solution while HC-HRS needs to iteratively determine the similarity between different channels making it considerably slower when compared to the NN solution.} Moreover, for SING, the choice of parameters $b_g$ and $r_g$ seems to harm the performance. We recall that both parameters are integers  thus are susceptible to the trade-off between $M$ and $G$. For instance, for $G=N=8$ and $M=12$, there exist only one viable option of $r_g$, i.e., $r_g = \lfloor M/G \rfloor = 1$.  Alternatively, we could select four ($\mathrm{mod}(M,G)$) groups to have $r_g = 2$, but this requires further processing on the choice of these groups. As a consequence, we obtain similar rates for $N=8$ users served with $M=8$ or $M=12$. Similar consequences are obtained for $N=12$. Moreover, for $G=N>M$, we have  $r_g =\mathrm{mod}(M/N) = 0$ what makes impossible to derive meaningful precoders Fig~\ref{fig:boxplots}(a)-(b). In contrast to that, the other three techniques, which consider clustering, do not suffer from this trade-off between $G, r_g$ and $M$. Instead, even for $N>M$ we still achieve reasonable spectral efficiency.


Finally, we analyse the capability of the shallow \gls{nn} to learn the grouping classification task as described above. To do so, we first analyse the accuracy of the network for class prediction. Recall that, here, a class represents a different clustering option.  Table~\ref{tab:summary_results} presents, in percentage, the results obtained by training different \gls{nn} according to the configuration parameters in Table~\ref{tab:simulation} for different number of users ($N$) and antennas in the BS ($M$).  The validation column contains the final classification accuracy in the validation dataset and indicates some learning capability in untrained data. During our experiments we noticed that different points of the same dendrogram might result in similar communication rates, i.e., there might exist different clustering options which achieve the same rate. Therefore, for the test dataset, we show the \textit{top-1}, \textit{top-3} and \textit{top-5} classification accuracy. Despite the fact that performance in \textit{top-1} accuracy might be considered poor, the \textit{top-5} results are, often, above $90\%$. Finally, the last row compares the communication rate decay (in \%) if using the \textit{top-1} option from the \gls{nn}. Results show that, except in the cases where $N>M$, on average, the rate drops $~2.5\%$ which is an acceptable loss when compared to the complexity of the original problem. Moreover, we can infer from these results that the \gls{nn} is capable of learning the maximum clustering option or clusters that approximate this option. In other words, it is capable to learn the relationship between different users directly from their channel matrices and cluster the users with a high degree of accuracy for most scenarios and finally achieve a rate comparable to more complicated similarity-based HC-HRS. 


\begin{table}[t!]
\centering
\caption{Summary of Results}
\begin{tabular}{|c|c|c|c|c|c|c|}
\hline
\textbf{N / M} &  \textbf{\begin{tabular}[c]{@{}c@{}}Validation\\ \textit{(top-1)}\end{tabular}} & \textbf{\begin{tabular}[c]{@{}c@{}}Test\\ \textit{ (top-1)} \end{tabular}} & \textbf{\begin{tabular}[c]{@{}c@{}}Test\\ \textit{(top-3)}\end{tabular}} & \textbf{\begin{tabular}[c]{@{}c@{}}Test\\ \textit{(top-5)}\end{tabular}} & \textbf{\begin{tabular}[c]{@{}c@{}}Test\\ Relative Rate\end{tabular}} \\ \hline
8 / 4  & 65.38\% & 65.37\%     & 85.22\%    & 90.48\%     & 94.12\%     \\ \hline
8 / 8   & 98.3\%     & 92.0\%    & 96.3\%    & 97.7\%    & 99.0\% \\ \hline
8 / 12  & 96.9\%     & 92.2\%    & 97.0\%    & 98.2\%    & 99.5\% \\ \hline
12 / 6    & 71.45\%   &35.6\%     & 65.62\%    & 77.75\%    & 89.99\%  \\ \hline
12 / 12  & 98.7\%     & 86.2\%    & 96.8\%    & 98.9\%    & 93.5\% \\ \hline
12 / 16  &  99.18\%   & 95.62\% & 98.32\% & 93.32\% & 99.77\%   \\ \hline
\end{tabular}
\label{tab:summary_results}
\vspace{-1.3\baselineskip}
\end{table}

\section{Conclusion }
In this work, we have proposed a \gls{nn} based clustering technique that learns and clusters users based on instantaneous noisy channel to maximise the rate achieved using Hierarchical Rate Splitting mechanism. The proposed technique is defined based on a shallow \gls{nn} architecture thereby making it extremely quick to learn and cluster the users based on the instantaneous noisy channel. The proposed technique is able to achieve a rate comparable with current works while being less complex compared to other techniques. Furthermore, this also helps to investigate further complex \gls{nn} structures such as Graph \gls{nn} which can learn covariances between different users to define clustering.

\bibliographystyle{IEEEtran}
\bibliography{./bibliography.bib}

\end{document}